# Benchmark Tests of Convolutional Neural Network and Graph Convolutional Network on HorovodRunner Enabled Spark Clusters

**Jing Pan, Wendao Liu, Jing Zhou\***

eHealth, Inc., Santa Clara, CA 95054
{jing.pan, wendao.liu, jing.zhou}@ehealth.com

## Abstract

The freedom of fast iterations of distributed deep learning tasks is crucial for smaller companies to gain competitive advantages and market shares from big tech giants. HorovodRunner brings this process to relatively accessible spark clusters. There have been, however, no benchmark tests on HorovodRunner *per se*, nor specifically graph convolutional network (GCN, hereafter), and very limited scalability benchmark tests on Horovod, the predecessor requiring custom built GPU clusters. For the first time, we show that Databricks' HorovodRunner achieves significant lift in scaling efficiency for the convolutional neural network (CNN, hereafter) based tasks on both GPU and CPU clusters, but not the original GCN task. We also implemented the Rectified Adam optimizer for the first time in HorovodRunner.

## Introduction

### Why HorovodRunner

It's a never-ending race for the industry to adopt distributed deep learning frameworks, and to prefer ones being easy to deploy and cloud-agnostic to minimize costs and risks in the ever-changing tide of the market. The Distributed TensorFlow comes with its strong tie to Google Cloud service and requires code modification with parameter servers, etc. Even though almost all deep learning frameworks have its own version of distributed training, similar obstacles motivate Uber to develop Horovod which implements data parallelism to take in programs written based on single machine deep learning libraries to do run distributed training fast (Sergeev and Balso, 2017). It's based on the MPI concepts of size, rank, local rank, allreduce, allgather, and broadcast (Uber, 2017; Sergeev and Del Balso, 2018). Although Horovod eases the code development process, it still requires companies to physically build a GPU cluster server on-premise, which presents another unbridgeable gulf for smaller companies to bridge the gap between them and big technology companies. For example, it's almost impossible for a smaller company in Silicon Valley to ensure power and cooling with California's wildfire prevention power shutdowns.

HorovodRunner is a general API to run distributed deep learning workload on Databricks Spark Cluster using Uber's Horovod framework (Databricks, 2019a). HorovodRunner and Horovod only support the four deep learning frameworks: TensorFlow, Keras, PyTorch, and Apache MXNet. HorovodRunner is also almost cloud-agnostic. On a Databricks Spark Cluster, the user can choose to use Azure or AWS for cloud service. It's currently proprietary to Databricks's Spark clusters. Yet it still poses a huge advance for smaller companies who are willing to pay for usage to finally enable distributed deep learning tasks, instead of investing a on premise GPU cluster or tying only to Google Cloud.

### *Status Quo* of HorovodRunner Benchmarks

There have been absolutely no benchmark tests on HorovodRunner, not even from Databricks internal testing. There are two benchmarks on Horovod. First, Uber's benchmarks show 90% scaling efficiency for Inception V3 and ResNet-101, and 68% scaling efficiency for VGG-16 with synthetic data set on GPU clusters (Uber, 2017). Second, on CPU clusters with custom data set and a very simple 1 hidden convolution layer NT3 model, the model.fit step achieved relatively linear scaling efficiency. With the bottleneck of data-loading time identified by their authors, the overall run time has yet shown evidence of scaling efficiency (Wu *et al.*, 2018). We suspect that in their study, both (1) the utilization of Horovod Timeline (Databricks, 2019b) and (2) utilization of a CPU cluster may contribute to the absence of the scaling efficiency. We optimize our experiments by avoiding Horovod Timeline and developing our version of the epoch time, and tested on both CPU and GPU clusters. The lack of overall scaling efficiency in the second study also implies that Uber had done a lot of engineering optimization in both their physical cluster set up and software architecture.

## Why GCN

CNN models were used in both the two existing benchmarks on Horovod and the Facebook data parallelization study (Goyal *et al.*, 2017) based on which Horovod was developed. GCN scales linearly with respect to the number of graph edges (Kipf and Welling, 2017). Our naïve motivation was that if the edges could be paralleled, it would have been a huge gain for the industry to utilize GCN outside of academia. In the industry, graph data can be so huge. Tencent in the past developed brutal force solution with the open-source Spark and Spark GraphX to handle some of their daily tasks like who is a friend of whom (Tencentbigdata, 2016). As people familiar with this matter know, the Spark community has made little progress towards GraphX over the years and the available graph computations are very limited. Neo4J, the market leader in the graph database (Rake and Baul, 2019), is not sharded. Each node needs to have an entire copy of graph and Neo4J's clustering mode only works for high query performance and its scaling options are complicated either with (1) Apache Mesos or (2) IBM POWER8 and CAPI flash systems together. Apache Giraph (completely) lost its momentum after 2016. Amazon Neptune comes with build-in auto-scaling capabilities. With the internet-scale of graph data, a distributed GCN implementation is necessary to put its application into production in the industry. When we see GCN was re-written with Keras model object (Kipf, 2018), we were determined to give it a try. We didn't try GCN's original TensorFlow implementation (Kipf and Welling, 2017) because we are not sure if their custom-built Model class can be pickled correctly with HorovodRunner.

## Methods and Results

All source code is uploaded to:
https://github.com/psychologyphd/horovodRunnerBenchMark.

## MNIST

At the very beginning of our study, we want to replicate the HorovodRunner's Keras MNIST example (Databricks, 2019c). We used the original MNIST dataset and the same model architecture as in Databricks's example. The task is to classify images of hand-written English digits to 10 labeled classes of digits. With a batch size of 128, initial learning rate of 0.1 on Adadelta optimizer, we trained 3 times on the data with 50 epochs each in each repetition on Databricks Spark Cluster with 6.1 ML runtime and consistently 8 C4.2xlarge (8 CPU per instance) AWS instances as workers. Single machine model used single machine Tensorflow Keras without HorovodRunner. The rest conditions used horovodRunner with different hvd.np. Note that hvd.np is setting of number of parallel processes to use for Horovod Job, if np<0 this will spawn np subprocesses on the drive node to run Horovod locally. If np>0 this will launch a Spark job with np tasks and run Horovod job on task nodes.

Because we are not using Horovod Timeline to get training time, we get time in the following two ways. First, we get wall-clock total time from right before calling HorovodRunner and after the HorovodRunner object's run method finished running, which include overhead in loading data, loading model, pickling functions, etc. Second, we get time to run each epoch (called epoch time) from the driver standard output. Every time a new line of the driver output is printed, we add a timestamp. In the driver standard output, it will print out "[1,x]<stdout>:Epoch y/z". x is the xth hvd.np, y is the yth epoch, and z is the total number of epochs. We record the timestamp t1 of the first time "Epoch y/z" shows up in the standard output, and the timestamp t2 of the first time "Epoch (y+1)/z", regardless of which process emits the output. The time difference, t2-t1, approximates the time spend for the epoch y to finish based on the assumption that only after all processes finish an epoch and average the weights, the next epoch can begin. For MNIST, we get wall-clock run time for the 3 repetitions and epoch times. We used number of images in the training set times number of repetitions and divided by total wall-clock time to get image per second.

The results are shown in Figure 1 and Table 1. It demonstrated scaling efficiency in a limited range of Horovod number of processes (aka, number of CPU/GPUs) when it is small or equal to 16. Hereafter, Horovod number of processes, Horovod size and hvd.np are used interchangeably. We did a one-way ANOVA on epoch time for all conditions. Results show the significant difference between all group to base group np=1 group.

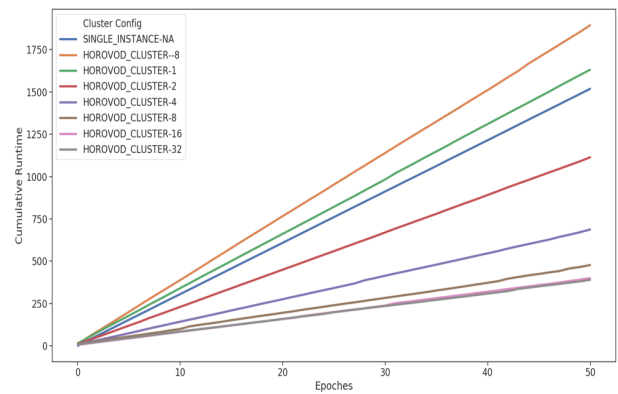

Figure 1: Scaling effect on a CPU spark cluster on MNIST. The figure shows scaling effect when np<=16 and reached asymptote when np>16.

Table 1: Time performance of HorovodRunner on MNIST dataset with CNN model.

| Machine | C4.2xlarge, 8 CPU | | | |
|---|---|---|---|---|
| Model | Databricks MNIST CNN | | | |
| Dataset | MNIST | | | |
| | Avg. epoch time (s) | Std. epoch time (s) | Tukey | Image/s |
| Single | 29.77 | *4.12* | *p*=0.001 | 183.94 |
| hvd.np=-8 | 31.98 | *2.54* | *p*=0.001 | 171.23 |
| hvd.np=1 | 37.13 | *4.32* | N.A. | 147.48 |
| hvd.np=2 | 21.84 | *1.65* | *p*=0.001 | 250.70 |
| hvd.np=4 | 13.48 | *1.22* | *p*=0.001 | 406.32 |
| hvd.np=8 | 9.37 | *1.71* | *p*=0.001 | 584.68 |
| hvd.np=16 | 7.85 | *1.17* | *p*=0.001 | 697.62 |
| hvd.np=32 | 7.65 | *1.00* | *p*=0.001 | 715.49 |
| ANOVA | *F*=1191.61, *d.f.*= (7, 400), *p*=2.766e-258 | | | N.A. |

### CNN Tasks

We used the Caltech101 dataset on VGG-16 and Inception V3 models. The task is to classify images of object into 101 labeled classes. Code was written with HorovodRunner and Keras; cluster was Databricks Spark Cluster with 6.1 ML GPU runtime and instance type was p2.8xlarge with 10 Gbps bandwidth. There are 2 slaves when hvd.np=16, and 4 slaves when hvd.np=32. There are 1 diver and 1 slave in the rest of the conditions. We used RMSprop with a learning rate of 0.001 and a batch size of 32. We demonstrated that Rectified Adam (Liu *et al.*, 2019) was able to be implemented with HorovodRunner as shown in our github code. Horovod and HorovodRunner have its own callback function that will handle learning rate change from all processes before broad cast the new changed learning rate to all processes. We only run 40 epochs once on the dataset for each condition. For more detail please see supplement part 1.

We found significant scaling efficiency for both models across our chosen hvd.np range as shown by significant two-way ANOVA and post hoc Tukey test. Consistent with Uber's results, we have higher scaling efficiency on Inception V3 (48.9%~79.7%) and lower scaling efficiency on VGG-16 (18.5%~49.0%). The results are shown in Figure 2 and Table 2.

### GCN

We are able to migrate the single machine Keras version of GCN code to HorovodRunner. The task is to classify scientific publications (nodes) in to one of the seven labeled classes. On the original Cora dataset, on a 1 driver and 2 slave CPU cluster with C4.xlarge instances, with 300 epochs, the training time (as displayed by Databricks notebook cell) on the single machine mode was 17.08s and with hvd.np=8 was 29.40s. We tried a GPU cluster, and upsampling cora dataset (not aiming at preserving the original graph properties, but simply at increasing number of nodes and edges). Nothing seems to show any scaling efficiency advantage of HorovodRunner. From hindsight, although the model object written in Keras can be pickled by HorovodRunner correctly, the adjacency matrix cannot take advantage of data parallelization based on which HorovodRunner improves scaling efficiency.

## Discussion

### HorovodRunner Scaling Efficiency on CNN

We confirmed the existence of the scaling efficiency of HorovodRunner for CNN tasks on both CPU and GPU clusters. Consistent with Uber's benchmarks, our scaling efficiency has the same trend that Inception V3's is higher than VGG-16. It may be explained by theoretical modeling of distributed data paralleled training reveals that at the batch size of either 10 or 60, the operational intensity of Inception V3 is lower than VGG16 (Castelló *et al.*, 2019). Uber used very high-end physical cluster setups, even in today's standard, which we can't afford. For more discussion on the cluster setup difference between Uber and us and how the difference will affect scaling efficiency, please refer to supplement part 2.

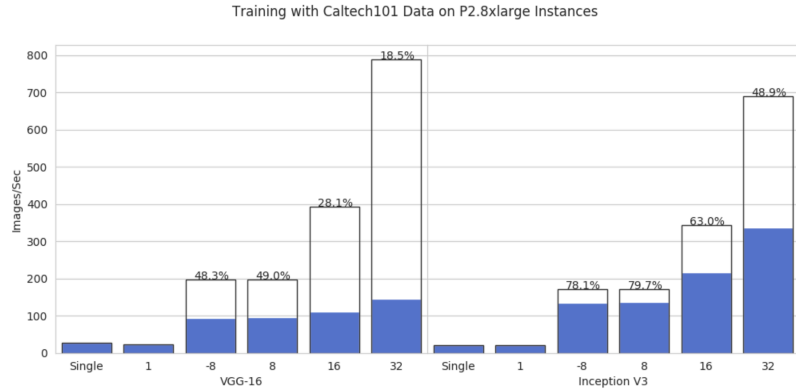

Figure 2: Scaling efficiency in VGG-16 and Inception V3 on GPU spark clusters. The scaling efficiency in Inception V3 is higher.

Table 2: Time performance of HorovodRunner on VGG-16 and Inception V3.

| Machine | P2.8xlarge, 8 GPU | | | | | | | |
|---|---|---|---|---|---|---|---|---|
| Dataset | Caltech101 | | | | | | | |
| Model | VGG-16 | | | | Inception V3 | | | |
| | Avg. epoch time (s) | Std. epoch time (s) | Tukey | Image/s | Avg. epoch time (s) | Std. epoch time (s) | Tukey | Image/s |
| Single | 237.71 | 47.09 | $p$=0.001 | 28.65 | 315.66 | 72.59 | $p$=0.9 | 21.57 |
| hvd.np=1 | 276.70 | 12.86 | N.A. | 24.62 | 316.29 | 69.48 | N.A. | 21.53 |
| hvd.np=-8 | 71.56 | 62.55 | $p$=0.001 | 95.19 | 49.61 | 7.96 | $p$=0.001 | 134.59 |
| hvd.np=8 | 70.59 | 11.96 | $p$=0.001 | 96.50 | 50.61 | 6.69 | $p$=0.001 | 137.31 |
| hvd.np=16 | 61.56 | 12.52 | $p$=0.001 | 110.65 | 31.39 | 7.90 | $p$=0.001 | 217.00 |
| hvd.np=32 | 46.83 | 12.85 | $p$=0.001 | 145.46 | 20.20 | 7.15 | $p$=0.001 | 337.29 |
| ANOVA | F=375.79, d.f.=(5, 234), p=2.603e-111 | | | N.A. | F=494.492, d.f.=(5, 234), p=3.66248e-124 | | | N.A. |

When hvd.np=1, HorovodRunner's performance is very similar to single machine Keras implementation, even if the single machine has 8 GPU/CPUs. That's because our single machine Keras implementation didn't implement any parallel multi GPU/CPU solution and the batch size can be fit into 1GPU/CPU. On a CPU cluster, when np=4 (on slave), the training is faster than a single machine implementation because of data parallelization. This scaling efficiency reaches asymptote quickly after when hvd.np>16. Even so HorovodRunner still possesses advantage over large batch size (current batch size*hvd.np) on single machine Keras implementation, because large batch size on single machine may affect model's accuracy, except done the manner that Horovod follows (Goyal *et al.*, 2017). Unlike the second benchmark study in the literature, we found a positive scaling effect. It confirms the effectiveness of our optimizations prior to our experiments. It's interesting for the researchers to find an optimal range of parameters/physical setting that balances the optimal GPU memory and computation power, the operational intensity and model's accuracy following the line of research on operational intensity (Castelló *et al.*, 2019).

**Prospect of GCN Application in the Industry**

Although direct use of HorovodRunner on Keras implementation of GCN doesn't yield performance gain with our initial attempt, GCN has potentials to accomplish HorvodRunner compatible implementations. Algorithms such as stochastic GCN (Chen Zhu and Song, 2017) doesn't need the entire neighbors of a node but only requires an arbitrary sample neighbors of a node. It hints possibilities for arbitrary data parallelization without partitioning the entire graph, which is crucial for HorovodRunner to gain performance.

For GCN to be widely used in the industry, in addition than graph database and model parallelization, it will be helpful to reduce the model's dependency on node labels. The original GCN paper (Kipf and Welling, 2017) still requires node label for training, and then predict labels for masked nodes. It's a common challenge for other types of models too, not specific to GCN. RNN based models were originally based on an explicit label, but moves to a more practical direction of getting rid of explicit label. It is true that RNN based models only deal with sequential relationships, and sequential relationships are much less intricate than the rich information contained in the edges of a graph. RNN based models, however, have achieved much less dependency on explicit labels. Perhaps the evolution of RNN based models would inspire the direction of GCN. We don't mean that GCN is less useful in the industry due to its dependency on labels in any way. In fact, given the potential of unsupervised learning, deep reinforcement learning on graphs might have the most use cases in the industry. In terms of GAN with GCN, the applications can be expanded beyond drawing graphs like drug molecules (You *et al.*, 2018). One possible application scenario is that when doing deep reinforcement learning on graphs, sometimes you don't have full graph information. If one can use any method, including GAN to generate the graph information, or Graphsage (Hamilton Ying and Leskovec, 2017) to add new node information, the reinforcement learning model could potentially gain performance.

## Acknowledgement

We thank Databricks engineers Parker Temple, Amy Wang, Vikas Yadav and anonymous engineers who we don't know their names for their help.

# Supplement

## 1. Methods and Results: CNN tasks.

We didn't use ResNet because it requires Tensorflow 2.0. We were able to (1) upgrade the cluster's 6.1 ML GPU runtime to use Tensorflow 2.0 and load Resnet model, (2) let HorovodRunner pickle Tensorflow 2.0 objects with "from tensorflow.compat.v1.keras import backend as K" statement. By the time of paper submission, we weren't able to get tensorflow 2 to config in HorovodRunner even with tf.compat.v1.ConfigProto(). Thus, we used Tensorflow 1.14 that comes with 6.1 ML GPU runtime for this study.

We tried all optimizers available in Keras, along with the very new Rectified Adam (Liu *et al.*, 2019) after installing PyPI package keras-retified-adam on the cluster. On the MNIST dataset and task, all optimizers work. On the 40 epoch Caltech 101 tasks, we initially found that except RMSprop, the rest of the optimizers would run several epochs but none of them can run to 40 epochs without an error complaining at the allReduce stage of HorovodRunner. So, we used RMSprop to report our results in our main paper. Later we discover that the allReduce stage error was caused by timeout of writing object to dbfs. With the help of Databricks engineering solution team, they found a fix of putting the following initiation script on the cluster and then Rectified Adam will be able to train successfully.

> dbutils.fs.put("tmp/tf_horovod/tf_dbfs_timeout_fix.sh",""" 
> #!/bin/bash
> fusermount -u /dbfs
> nohup /databricks/spark/scripts/fuse/goofys-dbr -f -o allow_other --file-mode=0777 --dir-mode=0777 --type-cache-ttl 0 --stat-cache-ttl 1s --http-timeout 5m /: /dbfs >& /databricks/data/logs/dbfs_fuse_stderr &""", True)

The Rectified Adam in HorovodRunner code is at:

> https://github.com/psychologyphd/horovodRunnerBenchMark/blob/master/np4_VGG_RAdam.html.

Batch size was chosen to be 32 for all Caltech 101 tasks for the following reasons: when batch size of 128 is chosen, it will be out of memory for GPU which is easy to understand. When a batch size of 64 is chosen, then hvd.np=32, we don't have enough testing data. 64*32=6048 images are needed for validation, but we only have 1695 images in the testing set when we split the Caltech101 data to train:test=4:1.

We load model from Keras applications, but we didn't use the model loaded directly in the get_model function in HorovodRunner directly, as in the HorovodRunner MNIST example would naturally lead into (Databricks, 2019). That is because each process will load the Keras applications model from github, and github will return the "too many request" error once your hvd.np becomes large (we begin to have this problem when hvd.np was set to 32 or almost always see this problem when it was set to 64). Our solution was to (2) load the model from Keras applications, (2) save it to master's drive as .h5 file, (3) then use dbfs.utiles to copy it to dbfs mounted on s3, and finally (4) in HorovodRunner, we used keras.model.load to load models from dbfs. We didn't use MLflow to serialize model to dbfs because MLflow has problem with HDF5 file on dbfs in Databricks 6.1 runtime.

We provide so much details on what doesn't work, which is intended to help our audience to avoid those troubles.

## 2. Discussion: Uber compared to our study

Uber's benchmark had higher image per second and scaling efficiency than our benchmark with the following uncompleted list of potential reasons. (1) HorovodRunner is another layer of complexity on Spark cluster that needs to communicate to driver, while Horovod's synchronous all reduce operation are communicated among peer training processes, without (a) parameter server(s) (as in distributed tensorflow) or a driver (as in Spark cluster). (2) Uber used higher network bandwidth, which is already identified as a factor affecting scaling efficiency (Castelló *et al.*, 2019). (3) Uber's data center, as far as the rumor goes in the industry, is an on premise distributed homebuilt system, while we used Databricks filesystem mounted on AWS S3. (4) Uber used different machine.

**Supplement references**